\documentclass[%
 aip,
 amsmath,amssymb,
 reprint, onecolumn
]{revtex4-1}
\usepackage{graphicx}
\usepackage{subfigure}
\DeclareGraphicsExtensions{.eps,.ps,.png,.pdf}
\usepackage{dcolumn}
\usepackage{bm}
\usepackage{url}
\newcommand{\Sotu}{SOTU}
\newcommand{\tcr}{$\mathit{TCR}$}

\usepackage[utf8]{inputenc}
\usepackage[T1]{fontenc}
\usepackage{mathptmx}
\usepackage{color}
\usepackage{dirtytalk}
\usepackage{etoolbox}

\long\def\omitit#1{}
\makeatletter
\def\@email#1#2{%
 \endgroup
 \patchcmd{\titleblock@produce}
  {\frontmatter@RRAPformat}
  {\frontmatter@RRAPformat{\produce@RRAP{*#1\href{mailto:#2}{#2}}}\frontmatter@RRAPformat}
  {}{}
}%

\newcommand{\review}[1]{\textcolor{black}{#1}}

\makeatother
\begin{document}

\preprint{AIP/123-QED}
\title[Timeline transformation via Surprisability]{Interpretable Transformation and Analysis of Timelines through Learning via Surprisability\footnote{Accepted for Publication in Chaos, May 2025}} 



\author{Osnat Mokryn}
 \affiliation{Information Systems, University of Haifa, Israel}
 \email{omokryn@is.haifa.ac.il}
 
\author{Teddy Lazebnik}
\affiliation{Cancer Biology, Cancer Institute, University College London, London, UK}%
\author{Hagit Ben Shoshan}
 \affiliation{Information Systems, University of Haifa, Israel}

\date{\today}

\begin{abstract}
The analysis of high-dimensional timeline data and the identification of outliers and anomalies is critical across diverse domains, including sensor readings, biological and medical data, historical records, and global statistics. However, conventional analysis techniques often struggle with challenges such as high dimensionality, complex distributions, and sparsity. These limitations hinder the ability to extract meaningful insights from complex temporal datasets, making it difficult to identify trending features, outliers, and anomalies effectively. Inspired by surprisability — a cognitive science concept describing how humans instinctively focus on unexpected deviations - we propose Learning via Surprisability (LvS), a novel approach for transforming high-dimensional timeline data. LvS quantifies and prioritizes anomalies in time-series data by formalizing deviations from expected behavior. LvS bridges cognitive theories of attention with computational methods, enabling the detection of anomalies and shifts in a way that preserves critical context, offering a new lens for interpreting complex datasets. We demonstrate the usefulness of LvS on three high-dimensional timeline use cases: a time series of sensor data, a global dataset of mortality causes over multiple years, and a textual corpus containing over two centuries of State of the Union Addresses by U.S. presidents. Our results show that the LvS transformation enables efficient and interpretable identification of outliers, anomalies, and the most variable features along the timeline.
\end{abstract}


\maketitle 
\textbf{The analysis of complex timeline data is essential in many fields. However, traditional analysis methods often struggle with the challenges posed by high-dimensional data, intricate patterns, and missing values operating on a trade-off of performance and explainability. This makes it difficult to identify important outliers, or anomalies. Inspired by the idea of \say{surprisability}, a concept from cognitive science that explains how humans naturally focus on unexpected changes, we introduce Learning via Surprisability (LvS). LvS analyzes high-dimensional timeline data by prioritizing unexpected deviations. It combines insights from cognitive science and computational techniques to detect anomalies in a way that maintains important context. We apply LvS to three real-world cases, showing that LvS can efficiently identify outliers, anomalies, and the most variable trends in these complex datasets, offering valuable insights for various fields.}

\section{Introduction}
High-dimensional timeline data appears in numerous domains, from biomedical research to economic modeling and global monitoring. Here, we use the term \say{timeline data} to describe temporally ordered datasets that may include event sequences, document corpora, or snapshot-based records (not limited to regularly sampled time series). These high-dimensional datasets track multiple, often interdependent, features evolving over time, making their analysis both computationally and conceptually challenging~\cite{box2015time}. Detecting significant changes, identifying anomalies, and extracting meaningful insights often become intractable due to issues such as non-Gaussian distributions, sparsity, and high feature interdependence ~\cite{donoho2000high}.

Traditional statistical and machine learning methods struggle to handle these complexities effectively, frequently requiring large labeled datasets or resorting to dimensionality reduction techniques that may obscure critical details~\cite{box2015time}.

Transformation methods, which are fundamental in mathematical analysis and data representation, offer a powerful approach to address these challenges by converting raw data into more structured and interpretable forms~\cite{n_intro_14,lacasa2008time}.
 These techniques enable the conversion of complex problems into more interpretable forms (formally, computationally appealing representation), revealing underlying structures and facilitating efficient computations. 
 
Feature-based transformations~\cite{fulcher2018feature} extract key numerical descriptors, improving efficiency in tasks like anomaly detection and forecasting, but their reliance on predefined metrics may overlook critical patterns. Graph-based transformations~\cite{lacasa2008time,gao2017complex} leverage network structures to model dependencies and nonlinear relationships, yet their effectiveness depends on the chosen graph construction, which can introduce biases and distortions. In addition, graph-based transformations can be computationally intensive, particularly for large-scale or real-time applications. 
While both approaches improve the representation of high-dimensional distributions, they also introduce abstraction layers that may obscure direct relationships and interpretability. 

A fundamental challenge in transformation methods is balancing interpretability with computational efficiency. Existing transformation methods define a Pareto front between their performance and explainability \cite{n_intro_17}. Classical techniques offer full analytical transparency but are often impractical for real-world data, while deep learning-based methods handle complexity but function as black boxes. Mokryn and Ben-Shoshan~\cite{mokryn2021domain} attempted to push this Pareto front with the Latent Personal Analysis (LPA) method, an information-theoretic transformation that provided a structured yet efficient transformation of data that proved efficient in authorship attribution in social media and impersonation detection. The method was also applied to immunology data in the body~\cite{alon2021using}. However, the LPA method and the other transformers are limited in their ability to capture time-series data. 
 
Building on these advancements, we propose Learning via Surprisability (LvS), a transformation designed to highlight unexpected deviations in high-dimensional timeline data. By quantifying surprisal, LvS provides a structured and interpretable transformation that enables effective anomaly detection without requiring predefined labels or heuristics. Unlike existing methods that either oversimplify data or obscure critical deviations,  LvS maintains computational efficiency while offering a transparent, human-aligned representation of unexpected events.

The LvS transformation constructs an expected distribution over the entire timeline and represents each time bin by its most surprising features, i.e., the features that deviate the most from this expected distribution. Intuitively, these are the features whose removal would most reduce the Jensen-Shannon divergence (JSD)~\cite{jensen99learning} between the time bin and the expected timeline distribution, thereby minimizing the Kullback-Leibler Divergence (KLD)~\cite{kullback1951information}.  Thus, the retained features are those that contribute most to distinguishing each time bin from the overall timeline.  

 The design of the LvS transformation is motivated by three key observations: \\
\review{\textbf{Missing Commonalities: }
Some of the most informative surprises in a time bin are features that are under-represented or entirely absent relative to their typical abundance over time. Because these features are common across the broader timeline, their absence in a specific bin is itself surprising. Such Missing Commonalities may arise from chance fluctuations or local selection processes that suppress otherwise prevalent patterns.} \\
\textbf{Per-Feature Decomposability of Divergence: } Both KLD and JSD decompose additively over features. This allows the LvS method to assign a divergence value to each individual feature, quantifying its contribution to the difference between a time bin and the expected distribution (the timeline center).\\
\textbf{Feature-Level Surprisability: } The most surprising features in a time bin are those with the highest per-feature divergence values. These features contribute most to the JSD and are retained in the Surprisability Profile to characterize what is unique or unexpected about the bin.

The proposed transformation aligns with the concept of surprisability, which quantifies the degree to which an event deviates from expected norms. Features that exhibit high surprisal are precisely those that cause the greatest divergence from the expected distribution. We hypothesize that this surprisability makes them the most informative in characterizing each time bin. By focusing on these features, LvS highlights the most unexpected yet significant variations in the data, offering an interpretable transformation that is both statistically grounded and cognitively aligned with how humans detect anomalies~\cite{levy2013memory}. 

Surprisability is fundamental in human comprehension and cognitive processes, where it is also tied with expectations. For example, in understanding a sentence, Surprisal Theory maintains that the processing difficulty is related to probabilistic expectations and is defined as the log of the inverse of the probability of the event. Given an observed new event, e.g., a new observed word in a string, the information value of the event is the reciprocal of its probability~\cite{levy2008expectation,norris2009putting}. 

In here, we present LvS, a novel transformation method, and demonstrate its power in identifying interpretable anomalies, outliers, and the most variable features along the timeline over three datasets, consisting of real-world sensor data, world main causes of death over a period of 30 years, and historical records.

The rest of this paper is organized as follows. Section \ref{sec:rw} provides an overview of transformation methods, and Anomaly detection methods. Section \ref{sec:model} formally outlines the proposed Learning via Surprisability method. Section~\ref{sec:res-ts} demonstrates LvS transformation over the two real-world time series datasets, and Section~\ref{sec:res-txt} demonstrates the LvS transformation over the timeline of textual data. Finally, section~\ref{sec:discussion} discusses the results in terms of possible applications and suggests future work. 

\section{Related Work}
\label{sec:rw}
In this section, we outline the applicative challenge LvS aims to tackle, as well as the computational methods in its base. Initially, we review the evolution of data transformation methods, providing context to the development of the field. Next, we review anomaly detections and their unique properties in high-dimensional and power-law distribution cases, followed by computational methods that use the distance between distributions for decision-making and the latent personal analysis algorithm.  

\subsection{Data transformation methods}
The study of mathematical transformations is rooted in the fundamental need to represent data in a way that is beneficial for other tasks, such as anomaly detection, clustering, data compression, and more. As such, the transformation processes in mathematics and data science have evolved significantly, transitioning from classical methods such as Taylor series and Fourier transforms to contemporary data-driven techniques like Principal Component Analysis (PCA)~\cite{n_intro_21} and autoencoders (AEs)~\cite{n_intro_22}.

The evolution from classical transformations to data-driven techniques reflects a broader trend in mathematics and data science, where the focus has shifted toward methods that can handle the complexities of modern data sets \cite{Rocha_2024}. Recent years have introduced two additional transformation approaches for time series, feature-based and graph network transformations~\cite{fulcher2018feature,lacasa2008time}. 

Feature-based time-series transformation involves extracting numerical descriptors that summarize key properties of temporal data, enabling tasks such as classification, clustering, and forecasting. Autocorrelation-based features capture dependencies between different time points by measuring how past values influence future ones. Entropy-based features measure the complexity or unpredictability of a time series. Other features characterize linear and non-linear dynamics or try to identify the generative structure of the data~\cite{fulcher2018feature}. 

The transformation of time series into graph networks leverages complex network theory and graph properties, thus enabling the analysis of the structural and topological features of the data~\cite{lacasa2008time,gao2017complex}. This transformation enables the study and detection of nonlinearity, chaos, extreme events, fluctuations, and temporal dynamics, including points of change~\cite{gao2017complex, zhang2017constructing, miller2020size}. 

In here, we suggest a timeline transformation algorithm that enables interpretable anomaly and outlier detection. The method borrows some of its concepts from Latent Personal Analysis, explained next.

\subsection{Anomaly and outlier detection in high dimensional data}
In domains like financial markets, network traffic, and sensor data, detecting anomalies is vital for monitoring, security, and accurate analysis. In these real-world domains, however, the data is high dimensional, making the task quite challenging~\cite{thudumu2020comprehensive}. 
In machine learning, anomaly detection (AD) is often framed as a binary classification task—distinguishing normal from anomalous data \cite{gornitz2013toward}. However, rare anomalies challenge traditional models tuned to the majority. This has led to specialized methods like Gaussian mixture models~\cite{li2016anomaly}, Mahalanobis distance~\cite{barnard2000detecting}, hypothesis testing~\cite{cohen2015active}, and convex hull-based techniques~\cite{costa2013partially}, all assuming a known data distribution. 
Recent methods span supervised, semi-supervised, and unsupervised approaches, with unsupervised AD favored for its low labeling demands~\cite{boukerche2020outlier}. However, many reduce dimensionality and lose interpretability. Timeline Analysis via Surprisability addresses this by enabling anomaly detection without sacrificing interpretability.

LvS is well-suited for high-dimensional data and performs especially effectively on long-tailed distributions. Practically, long tail and power-law distributions are prevalent in nature and empirical data~\cite{newman2005power}. 
The complexity of high-dimensional spaces often leads to difficulties in identifying anomalies due to the \say{curse of dimensionality} \cite{verleysen2005curse}. This phenomenon is exacerbated in multivariate contexts, where the interaction between multiple dimensions can obscure the identification of outliers \cite{rw_3}.

Various methodologies have been proposed to address these challenges \cite{teddy_ch_volume}, among them,  For example, leveraging AEs to transform high-dimensional classification datasets into formats suitable for anomaly detection~\cite{rw_4}, and using dimensionality reduction techniques~\cite{rw_5}. However, as dimensionality increases, dimensionality reduction techniques might obscure critical information~\cite{chari2023specious}.


Currently, there are widely adopted unsupervised anomaly and outlier detection methods considered state-of-the-art. Among them are the following. First, the Isolation Forest~\cite{cheng2019outlier} algorithm isolates individual data points through recursive partitioning of the data space, identifying outliers based on the speed (i.e., number of partitions) at which they can be separated. It constructs an ensemble of isolation trees, where data points that are isolated with shorter average path lengths are flagged as anomalies. This method relies solely on the ordinal ranking of each variable, disregarding distances and inter-variable relationships. Thus, its effectiveness may be diminished in situations where these factors are critical, like in high-dimensional spaces. Second, \textit{Single-Class SVM}~\cite{oza2018one} detects anomalies by learning a boundary around normal data. Trained on data from a single class, these methods assume points outside the learned boundary are anomalous. \textit{Single-Class SVMs} construct a hyperplane that maximally separates normal data from the origin. Third, \textit{Local Outlier Factor} (LOF)~\cite{alghushairy2020review} measures the local density deviation of a data point relative to its neighbors, identifying outliers in areas of significantly lower density. This method detects abnormality without considering the global structure of the data and, therefore, will under-perform in cases where the context of the data can shed more light on its place in a distribution represented by a more computationally meaningful space.

\section{Timeline transformation using Learning Via Surprisability}
\label{sec:model}

\subsection{Motivation}

This work builds on the observation that a timeline — composed of a sequence of time bins — can be transformed to highlight the most surprising elements in each bin relative to an overall expected distribution.
 
LvS handles high-dimensional, long-tailed data effectively by identifying surprising features, including missing commonalities often overlooked by other methods.

\subsection{LvS transformation algorithm}
Let $\mathcal{T}$ denote a time axis (or timeline). We assume here that the timeline period is known and that $N$ is the total number of bins in the timeline  $\mathcal{T}$\footnote{In later works, we intend to expand to a streaming setting.}. 
We define $T_\imath, \imath \in [1..N]$, as the high-dimensional probability distribution of elements at time bin $\imath$. Each $T_\imath$ distribution comprises a set of elements $E_\imath \subseteq \Omega $, where $\Omega$ is the set of all elements that appear in any of the bins in  $\mathcal{T}$. In our case, we use the terms element and feature interchangeably. 

We characterize a time axis $\mathcal{T}$ that consists of $N$ high dimensional probability distributions using a Timeline Center Representation for the timeline, and a Surprisability Profile for each of the time bins in the timeline, as follows. 

The LvS transformation consists of three main steps. First,  it constructs a timeline center representation (\tcr) which is a global expectation across all time bins. Second, it computes a Surprisability Profile for each time bin, which highlights how each bin deviates from the \tcr. Lastly, filtering each SP to retain only the most surprising elements above a threshold, it yields an interpretable, sparse representation. Together, these steps provide both a scalar divergence from the norm (LvS Divergence) and a compact interpretable vector (LvS SP).

\subsubsection{Creating a Timeline Center Representation} 
Pivotal to the proposed method is the creation of a timeline’s information-theoretic center representation, which captures the average distribution of features across all time bins. This center serves as the expected baseline against which each time bin is compared, enabling the identification of features that are unusually over- or under-represented. The Timeline Center Representation ($\mathit{TCR}$) is a distribution that consists of the mean value for each element $e \in \Omega$, calculated across all time bins. An element $e$ can be in an arbitrary $k$  number of time bins, $1 \leq k \leq N$, where N is the total number of time bins in  $\mathcal{T}$. We determine the  $\mathit{TCR}$ by calculating the average value for each element across all time bins in the timeline. If an element is missing in a specific time bin distribution $T_\imath$, it is calculated as zero, as if this time bin added zero to the mean. 

To create the global expectation, we compute the average relative frequency of each element across all bins. Elements missing in a bin are treated as zero, contributing nothing to the average. The Timeline Center Representation ($\mathit{TCR}$) is then calculated as follows:

\begin{widetext}
\begin{equation}
\mathit{TCR}(j) = \frac{1}{N_j} \sum_{\imath=1}^N T_\imath(j) \cdot \mathbb{I}(\text{element } j \text{ exists in } T_\imath)
\end{equation}
\end{widetext}
\begin{widetext}
\begin{equation}
    \mathit{TCR} = \left\{ \mathit{TCR}(j), j \in \Omega  \right\}
    \label{eq:tcr}
\end{equation}
\end{widetext}
Sorting the $\mathit{TCR}$ in decreasing order of mean relative frequencies will yield a ranked list of the elements, according to their {\em relative prominence} in the timeline. 

\subsubsection{Creating a time bin Surprisability Profile: } For a given time bin distribution $T_\imath$, the Surprisability Profile (SP) identifies the elements $e \in \Omega $  that most strongly deviate from the expected distribution represented by the timeline center (\tcr). To quantify this deviation, we calculate the JSD between $T_\imath$ and \tcr.  The JSD is a symmetric measure based on KLD. Unlike KLD, JSD is bounded and symmetric, making it suitable for comparing distributions that may have non-overlapping support. It allows us to decompose the divergence into per-feature contributions. These contributions form the basis of the SP, with each element assigned a weight indicating its surprisal magnitude and a sign denoting whether it is over- or under-represented.

\begin{widetext}
\begin{equation}
\textit{JSD}(\mathit{TCR}, T_\imath) = \frac{1}{2}\textit{KLD}(\mathit{TCR}, M_\imath) + \frac{1}{2}\textit{KLD}(T_\imath, M_\imath)
\label{jsd}
\end{equation}
\end{widetext}
\textit{where}
\begin{widetext}
\begin{equation}
M_\imath = \frac{1}{2}(\mathit{TCR} + T_\imath)
\end{equation}
\end{widetext}
and \(\textit{KLD}\) denotes the \textit{asymmetric} Kullback-Leibler Divergence.

We then identify the time bin's SP as follows. 
For each time bin \(\imath\), the JSD calculated between the center and the time bin, created a per-element divergence between the value of the element in the time bin (zero if missing) and their mean value in the \(\mathit{TCR}\), which is a non-zero value. Consequently, the process generates \(\mathit{SP}_\imath\), a local vector representation of the time bin's distribution, defined as:

\begin{widetext}
\begin{equation}
\forall j \in \Omega, \quad \mathit{SP}_\imath(j) = \left(\frac{1}{2} \mathit{TCR}(j) \log_2\left(\frac{M_\imath(j)}{\mathit{TCR}(j)}\right) + \frac{1}{2} T_\imath(j)\log_2\left(\frac{M_\imath(j)}{T_\imath(j)}\right)\right)\cdot\mathbb{I}(\text{element } j \text{ exists in } T_\imath)
\label{eq:sp-init}
\end{equation}
\end{widetext}

Let $D_\imath$ denote the divergence of time Bin $T_\imath$ from the $\mathit{TCR}$, given by: 
\begin{widetext}
\begin{equation}
        D_\imath = \sum_{j=1}^\Omega  |SP_\imath(j)|\cdot\mathbb{I}(\text{element } j \text{ exists in } T_\imath)
\label{eq:lvsfd}
\end{equation}
\end{widetext}
We assign each element in \(\mathit{SP}_\imath\) a weight corresponding to its contribution to the JSD. We then assign a sign to this weight: positive if the feature is overused in the time bin compared to the timeline center, and negative if it is underused. This results in a signed surprisal value that captures both the magnitude and direction of deviation.
\begin{widetext}
\begin{equation}
\forall j \in \Omega, \quad \mathit{SP}_\imath(j) = \mathbb{I}(\text{element } j \text{ exists in } T_\imath)\cdot\begin{cases}
\mathit{SP}_\imath(j) & \text{if } M_\imath(j) \geq \mathit{TCR}(j) \\
-\mathit{SP}_\imath(j) & \text{if } M_\imath(j) < \mathit{TCR}(j)
\end{cases}
\label{eq:sp-sign}
\end{equation}
\end{widetext}

The next step is to determine the subset of elements in \(\mathit{SP}_\imath\) for which the calculated value is sufficiently surprising. Only elements that exceed a predefined surprisability threshold \(\theta\) will be retained to define \(\mathit{SP}\). Formally, this subset is given by:

\begin{widetext}
\begin{equation}
\mathit{SP}_\imath^\theta = \left\{ j \in \mathit{SP}_\imath \mid |\mathit{SP}_\imath(j)| > \theta \right\}
\label{eq:sp-theta}
\end{equation}
\end{widetext}

where \(\theta\) denotes the surprisability threshold. The elements in \(\mathit{SP}_\imath^\theta\) represent those considered significantly surprising within the time bin $\imath$. In our implementation, the threshold $\theta$ is selected to retain elements that contribute substantially to the total JSD between each time bin and the timeline center. Specifically, we use a fixed surprisal value as a cut-off, chosen empirically to ensure that all elements with a meaningful influence on the divergence are preserved, while excluding minor contributors and noise. This results in compact, interpretable profiles without sacrificing the most informative deviations. Although this threshold is currently static, it can be adapted based on dataset characteristics or statistical significance.

LvS transformation identifies each time bin $T_\imath$ by two representations.  The first is its JSD, and the second is its SP, as follows. \\
\textbf{LvS Divergence: }  $D_\imath$ is a measure of how different is $T_\imath$ from the dataset when the $\mathit{TCR}$ is expected.  It is calculated as the sum of the absolute values of the JSD differences of $T_\imath$'s elements from their corresponding elements in the $\mathit{TCR}$'s, where timeline prominent elements that are missing from $T_\imath$ contribute to this divergence.  $D_\imath$ is defined in Eq~\ref{eq:lvsfd}. \\
\textbf{LvS SP: } $SP_\imath^\theta$ is a low dimensional interpretable presentation of $T_\imath$ that includes the most surprising elements in $T_\imath$ when $\mathit{TCR}$ is expected. The threshold-based SP, $SP_\imath^\theta$, is defined in Eq~\ref{eq:sp-theta}.

The LvS method is grounded in fundamental concepts from information theory. At its core is the notion of self-information or surprisal, defined as -log[p(x)], which underpins Shannon entropy and cross-entropy. In our approach, each time bin is compared to the timeline’s expected behavior, represented by the center distribution (\tcr), using the Jensen-Shannon divergence (JSD). Because JSD decomposes into per-feature contributions, LvS captures the surprisal of individual features, resulting in interpretable Surprisability Profiles (SPs) that highlight which features are over- or under-represented. In this way, LvS connects directly to the machinery of information theory: entropy, cross-entropy, and divergence are not only part of the underlying computation, but also key to interpreting the resulting profiles in a mathematically grounded and explainable way.

\section{Timeline analysis with LvS: timeseries}
\label{sec:res-ts}
 LvS's approach of finding the uniqueness of each time bin compared with the timeline while accounting for latent elements along the timeline enables timeline and timeseries analysis,  as is demonstrated in the following. 

\subsubsection{Analyzing SWaT: Multimodal Time Series Anomaly (Attack) Prediction}
\label{sec:swat}

To empirically evaluate LvS's performance with respect to the current state-of-the-art results, we adopted the popular \textit{SWAT dataset}~\cite{Yuan2019,bozdal2024comparative,bozdal2023security}, a real-world data from the Secure Water Treatment plant testbed, a testbed built at the Singapore University of Technology and Design~\cite{goh2017dataset}, and publicly available (\url{https://itrust.sutd.edu.sg/itrust-labs_datasets/dataset_info/}). The SWAT dataset describes a computational system's state over time with 450 thousand records and 51 features for each record. Out of these, around 12\% of the records are manually tagged as cyber security attack attempts. 
 
\begin{figure}[!ht]
     \centering
     \includegraphics[width=0.5\linewidth]{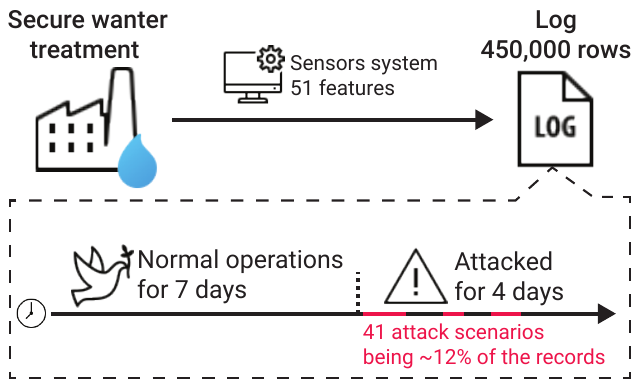}
     \caption{Schematic illustration of the SWaT dataset: a six-stage water treatment process with sensor and actuator data collected under normal and attack conditions. The figure highlights the temporal structure of the dataset and the distribution of attack labels across the timeline.}
     \label{fig:swat}
 \end{figure}

The SWaT dataset consists of sensor and actuator data collected from a fully operational six-stage water treatment process. It contains two main types of data: normal operation and under attack. For the normal operation case, data was collected over 7 days, where the system was operating under standard conditions without cyber or physical attacks. For the attack case, data was collected over 4 days, during which a total of 41 carefully designed cyber-physical attack scenarios were executed. These attacks include sensor spoofing, actuator manipulation, and communication disruptions, targeting different stages of the water treatment process.  Each row in the dataset is a timestamped record containing 51 features, including sensor readings (e.g., flow rates, water levels, conductivity), actuator states (e.g., pump status, valve positions), and a label indicating whether the system was under attack or in normal operation. Importantly, as the data is designed for research, the manually tagged cyber security attack attempts capture all the cyber security attacks in the data and are fully accurate. To help clarify the structure of the dataset and the distribution of attack events over time, Figure~\ref{fig:swat} provides a schematic overview of the SWaT system and its data collection process, highlighting the days of normal operation and those during which attacks were conducted.

First, we predict the anomalies using the \textbf{LvS Divergence} representation and compare the performance to state of the art unsupervised algorithms. Our hypothesis is that if the majority of the time bins register the system state during normal operation, the JSD of time bins registering the system state when under attack would be larger, and hence their LvS Divergence would be bigger. We use the LvS transformation over the SWaT data, and train the Isolation Forest, LOF, One-class SVM, Long short term memory (LSTM)~\cite{malhotra2015long}, and fully connected AE models on attack detection. \review{We used grid search to tune all models, with the full hyperparameter search spaces detailed for each algorithm, ensuring fair comparison and alignment with prior studies (See Apendix A). All experiments were run on a clean GCP instance (n1-standard-2, Ubuntu 18.04, Python 3.9), using scikit-learn for classical models and PyTorch for neural networks, ensuring reproducibility with standard libraries and configurations.} In order to avoid differences in decision threshold optimization, the optimal recall-precision threshold is used for all algorithms.  Table \ref{tab:anomaly_comparison} shows the results of this analysis in terms of the model's area under the receiver operating characteristic curve (AUC), recall (sensitivity), precision, \(F_1\) score, and accuracy. Notably, LvS obtains the highest AUC value (82.3\%) followed by a large margin by Isolation Forest with 54.5\%. For precision, the Isolation Forest obtains the highest precision (97.5\%) while achieving a recall of 8.0\% while one-class SVM received a precision of 91.3\% and recall of 76.0\% which also results in the highest F1 score of 83.0\% followed by LvS. Taken jointly, while the one-class SVM is able to differentiate between the anomaly and regular records the best, its decision threshold, as indicated by its low AUC score (9.1\%) is hard to find and therefore deeming the results unstable. \review{Average inference time per sample (over 100,000 samples) ranged from 0.0002 seconds for SVM to 0.0078 seconds for AE, with LvS at 0.0013 seconds—comparable to classical methods like Isolation Forest (0.0009) and LOF (0.0011).}

 \begin{table*}[!ht]
    \centering
    \begin{tabular}{lcccccc} 
        \hline \hline
        \textbf{Metric} & \textbf{LvS} & \textbf{Isolation Forest} & \textbf{LOF} & \textbf{One-Class SVM} & \textbf{LSTM} & \textbf{Fully connected AE} \\ 
        \hline \hline
        AUC & \textbf{82.3\%} & 54.5\% & 50.3\% & 9.1\% & 78.5\% & 71.0\% \\  
        Precision & 69.9\% & \textbf{97.5\%} & 22.9\% & 91.30\% & 75.2\% & 65.0\%  \\
        Recall & 69.9\% & 8.0\% & 3.4\% & \textbf{76.0\%} & 72.0\% & 60.5\% \\ 
        F1 Score & 69.9\% & 14.8\% & 5.8\% & \textbf{83.0\%} & 73.6\% & 62.7\% \\ 
        Accuracy & 92.7\% & 88.8\% & 93.4\% & \textbf{96.21\%} & 94.5\% & 93.8\% \\ 
        \hline  \hline
    \end{tabular}
    \caption{Performance comparison of different anomaly detection methods, the best value of each metric is highlighted in bold font.}
    \label{tab:anomaly_comparison}  
\end{table*}

Moreover, as these results are obtained for the optimal decision threshold, they may not reflect the performance of LvS in applicative settings. As such, we conducted a sensitivity analysis around the optimal decision threshold, analyzing the performance of LvS for each case. Table \ref{tab:performance_metrics} presents the results of this analysis, where a distance of ten percent from the optimal decision threshold results in a drop of two percent in both recall and precision. The change in the AUC and accuracy is even smaller, with less than a one percent decline. 

\begin{table}[!ht]
    \centering
    \begin{tabular}{lccccc}
        \hline \hline
        \textbf{Metric} & \textbf{-10\%} & \textbf{-5\%} & \textbf{0\%} & \textbf{5\%} & \textbf{10\%} \\
        \hline \hline
        AUC        & 81.8\% & 82.4\% & 82.9\% & 83.3\% & 83.6\% \\
        F1 Score   & 67.9\% & 69.0\% & 69.9\% & 70.8\% & 71.3\% \\
        Precision  & 67.9\% & 69.0\% & 69.9\% & 70.8\% & 71.3\% \\
        Recall     & 67.9\% & 69.0\% & 69.9\% & 70.8\% & 71.3\% \\
        Accuracy   & 92.3\% & 92.5\% & 92.7\% & 92.9\% & 92.9\% \\
        \hline \hline
    \end{tabular}
    \caption{Decision threshold sensitivity analysis for LvS on the SWAT dataset.}
    \label{tab:performance_metrics}
\end{table}

We continue here and evaluate the performance of our transformation in identifying anomalies using the \textbf{LvS SP} representation. 
The hypothesis is that the SP of time bins during normal operation will resemble each other more than the SP of time bins during attacks. 
To prepare the LvS data for K-means clustering, a scaling procedure was applied to eliminate negative values. The preprocessing consisted of two steps: first, vector elements were multiplied by 10, and then a constant of 0.4 was added to ensure non-negativity. The scaled LvS vectors were then processed with PCA before clustering using K-means with 2, 3, and 5 clusters. The clustering performance was assessed by comparing the generated cluster assignments with the predefined 'Attack' labels. Table \ref{tab:confusion_matrix_metrics} presents the confusion matrix and performance metrics of LvS. 

\begin{table}[h]
    \centering
    \caption{Confusion matrix and performance metrics}
    \begin{tabular}{lccc}
        \hline
        \textbf{True Label} & \textbf{Cluster 0 } & \textbf{Cluster 1 } & \textbf{Total (Actual)} \\
        \hline
        Normal Operation & 395,187 & 111 & 395,298 \\
        Attack & 22,661 & 31,960 & 54,621 \\
        \hline
        \textbf{Total (Predicted)} & 417,848 & 32071 &  \\
        \hline
    \end{tabular}
    \vspace{0.5cm}

    \begin{tabular}{lcc}
        \hline
        \textbf{Metric} & \textbf{Cluster 0} & \textbf{Cluster 1} \\
        \hline
        Precision & 0.9458 & 0.9965 \\
        Recall & 0.9997 & 0.5851 \\
        F1 Score & 0.9720 & 0.7373 \\
        \hline
        \textbf{Overall F1 Score} & \multicolumn{2}{c}{0.8547} \\
        \hline
    \end{tabular}

    \label{tab:confusion_matrix_metrics}
\end{table}
The clustering results confirm the effectiveness of the LvS SP representation in distinguishing normal operation from attacks. Cluster 0 captures most normal instances with 0.9997 recall, while Cluster 1 identifies attacks with 0.9965 precision, though its lower recall (0.5851) indicates missed detections. The overall F1 score of 0.8547 reflects strong but imbalanced classification. Clustering with 3 and 5 clusters yielded poor performance and was therefore omitted.

\subsubsection{Interpretable Outlier and Anomaly Detection}
\label{sec:oad}
To visually demonstrate the interpretability and outlier detection capabilities of LvS, we apply it to a smaller dataset: 30 years of global Causes of Death (CoD) data spanning 1990–2019~\cite{murray2022global}. This dataset reports the number of deaths attributed to the top 30 causes worldwide for each year. Importantly, deaths are categorized by underlying causes rather than contributing risk factors (e.g., smoking)~\cite{murray2022global}. Since the dataset ends in 2019, it does not include COVID-19–related mortality.

To prepare this dataset for LvS transformation, we created a timeline of yearly probability distributions (time bins), where each year is represented as a vector of the 30 causes. For each year, we obtained the global population and computed the proportion of deaths from each cause relative to the total population. This yielded one probability distribution per year, with vector features corresponding to specific causes of death.

We begin with a visual, interpretable anomaly detection using LvS, followed by a numerical analysis. Specifically, we construct two distance matrices: $A$ and $ASP$. Matrix $A$ captures the pairwise $L_1$ (Manhattan) distances between the original yearly probability distribution vectors of causes of death, while $ASP$ captures the distances between the corresponding yearly SPs ($SP^\theta$) generated by LvS. For each pair of years $i$ and $j$, the matrix entries are defined as:
\begin{widetext}
\[
A_{ij}=\sum_{f=1}^{30}|V_i(f)-V_j(f)|, \quad ASP_{ij}=\sum_{f=1}^{30}|SP^\theta_i(f)-SP^\theta_j(f)| \quad \text{for } i,j \in [1990, 2019] 
\]
\end{widetext}

These matrices allow us to compare how years differ in terms of their raw distribution of causes of death (via $A$) and how they differ in terms of their most surprising or anomalous features (via $ASP$), as identified by the LvS transformation.
\begin{figure*}[t]
{\subfigure[Distance measure across all time bins' distributions in world mortality data of 30 years, original data]{\includegraphics[width=0.49\textwidth]{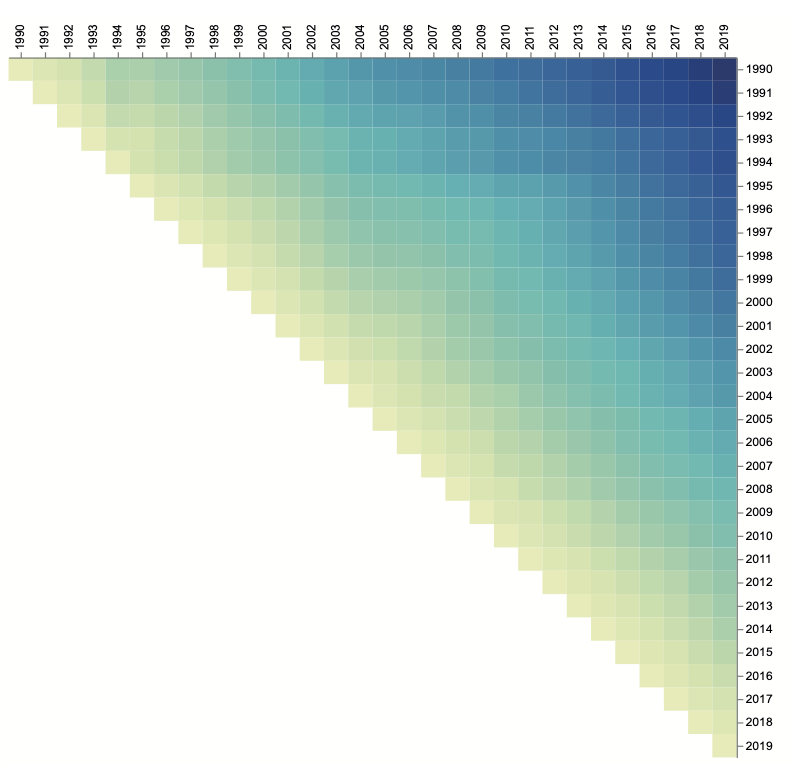}}\label{fig:cos-original}}
{\subfigure[Distance measure across all time bins' SPs, identifying points of change in world mortality reasons]{\includegraphics[width=0.49\textwidth]{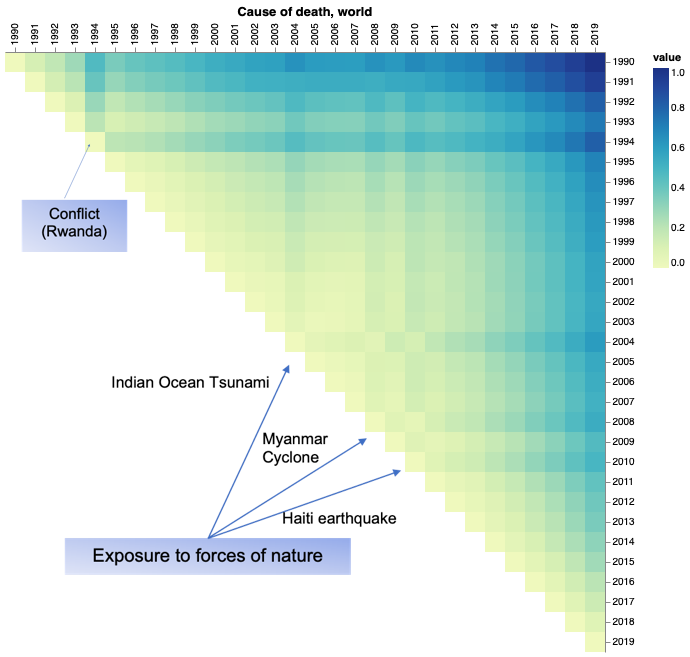}}\label{fig:cos-SP}}
\caption{Analysis of causes of death over 30 years: (a) A comparison of the original vectors representing causes of death over a 30-year period. Each square corresponds to the comparison of probability distribution vectors between two years, with lighter colors indicating greater similarity. The figure clearly shows that any two consecutive years exhibit a high degree of similarity in the underlying causes of death worldwide. (b) A comparison of the Surprisal Profile vectors created by LvS of the causes of death over a 30-year period. Each square corresponds to the comparison of the SP probability distribution vectors between two years, with lighter colors indicating greater similarity. Notable anomalies are clear in the years 1994, 2004, 2008, and 2010. The values within the SP vectors enable interpretation of the most surprising elements causing the anomaly in these years. Values were normalized for easier reading.}
\label{fig:cod-sbs}
\end{figure*}

Figure~\ref{fig:cod-sbs} presents a visual side-by-side comparison of $A$ and $ASP$, that is, the cross-comparison results between the yearly probability distribution vectors (Figure~\ref{fig:cod-sbs} (a)) and the yearly SPs (Figure~\ref{fig:cod-sbs} (b)). Each square represents a comparison between two years, as indicated by the corresponding row and column, with lighter colors denoting greater similarity. The distance matrices were normalized before applying the coloring. 

Figure~\ref{fig:cod-sbs} (a) clearly illustrates that consecutive years exhibit a high degree of similarity in the underlying causes of death worldwide. There is a high correlation between subsequent time bins, and the divergence of change is subtle and incremental. Death rates associated with diseases, illnesses, and other health factors, which account to the majority of world-wide deaths, generally evolve gradually over time. While they may fluctuate slightly from year to year as part of broader trends, significant shifts are uncommon, disregarding natural disasters and terrorism-related deaths. 

Indeed, when comparing the SP vectors across the years, as shown in Figure~\ref{fig:cod-sbs} (b), we obtain a visually interpretable method for outlier detection, where significant change points become evident. This allows for the identification of specific moments in time when natural disasters and terrorism-related deaths peak. The SP also indicates the most surprising feature, as depicted in Figure~\ref{fig:cod-sbs} (b). These findings are the same as the in-depth analysis in~\cite{murray2022global}. A change was evident in 1994, and the corresponding most surprising element in the SP is \textit{conflict}. It is attributed to the genocide in Rwanda in 1994, that ``stands out for its very high death-toll''~\cite{murray2022global}. The years 2004, 2008, and 2010 all are anomalous. In these years, the most surprising feature is  \textit{exposure to forces of nature}, corresponding to the 2004 Indian Ocean earthquake and tsunami; Cyclone Nargis which struck Myanmar in 2008; and the 2010 Port-au-Prince earthquake in Haiti~\cite{murray2022global}. 

In summary, LvS transformation enables interpretable anomaly detection. 

\subsubsection{A Measure of Dynamics: Identifying the Most Variable Features}
We define highly variable features as those that exhibit significant fluctuations across the timeline, reflecting substantial temporal changes in the high-dimensional distributions. A key measure of timeline dynamics, where each time bin represents a high-dimensional distribution, is the extent to which individual features vary over time. Unlike traditional approaches that primarily capture trends, our method, LvS, uniquely identifies time bins in which certain features are missing or underutilized. This provides a more comprehensive view of feature dynamics, ensuring that critical variations are not overlooked. Additionally, LvS effectively pinpoints the features that undergo the most significant changes along the timeline, offering deeper insights into temporal variability.
\begin{figure*}[!ht]
 \includegraphics[width=.8\linewidth]{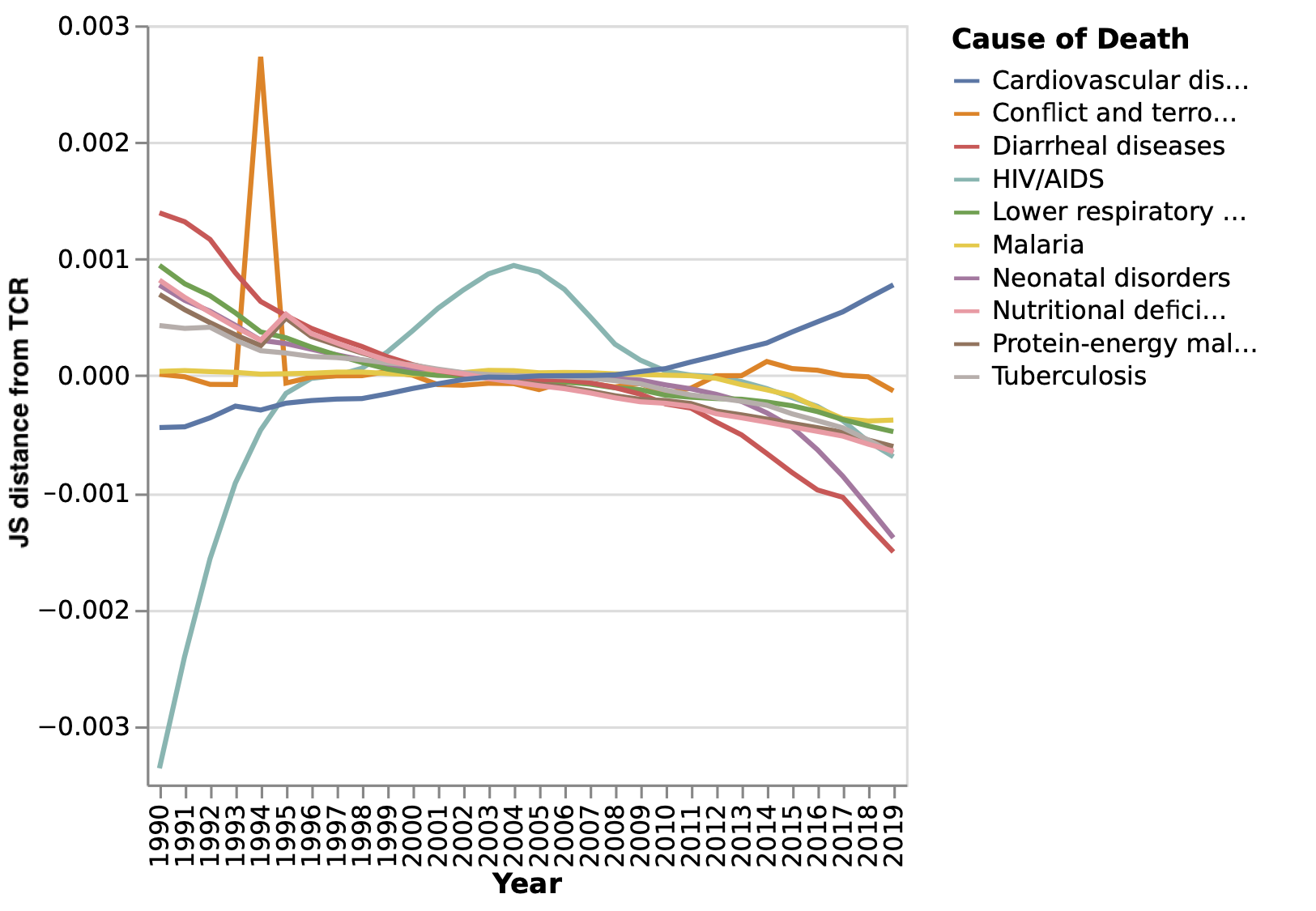} 
 \caption{Divergence from the \tcr~of the 10 most highly variable (fluctuating along the timeline) causes of death in the CoD dataset. In 1994, for example, there is a spike in deaths from conflict, making it one of the primary reasons this year stands out significantly from the surrounding years, as seen in Figure~\ref{fig:cod-sbs}.}
 \label{fig:top-10}
\end{figure*}
 To identify the terms that were the most highly variable across the timeline, we looked for those that exhibited the greatest fluctuations in their relative importance compared to their mean value in the \tcr. For each time bin $\imath$,  Equations~\ref{eq:sp-init} and ~\ref{eq:sp-sign} identify the JSD of each element (feature) from their value in the \tcr. 

For each element \( e \in \Omega \), we define its accumulated divergence from the \tcr~as:

\begin{widetext}
\begin{equation}
    D_{js} (k) = \sum_{\imath \in N} |SP_\imath (k)|
\label{eq:fluct1}
\end{equation}
\end{widetext}

Sorting \( D_{js} \) in descending order provides a ranking of feature variability:

\begin{widetext}
\begin{equation}
    D_{sp}^{\downarrow} = \text{sort}(D_{sp}, \text{descending})
\label{eq:sortedD}
\end{equation}
\end{widetext}

where \( D_{sp}^{\downarrow} \) represents the sorted version of \( D_{sp} \), arranged in decreasing order:

\begin{widetext}
\begin{equation}
    D_{sp}^{\downarrow} (1) \geq D_{sp}^{\downarrow} (2) \geq \dots \geq D_{sp}^{\downarrow} (|\Omega|)
\label{eq:sorted_order}
\end{equation}
\end{widetext}

This ordering ensures that the most highly variable features along the timeline appear at the top.

Figure~\ref{fig:top-10} presents the top 10 most highly variable causes of death over the 30-year period. Interestingly, while HIV was a major topic of public discussion during the 1990s, its peak in mortality occurred in 2004 and 2005. Additionally, notable is the growing number of deaths attributed to cardiovascular disease over the years, while deaths from diarrheal diseases show a steady decline during this period. 

\section{Analysis of timelines of textual data: the State of the Union Addresses 1790--2022 dataset}
\label{sec:res-txt}
Here, we use the State of the Union dataset~\cite{stou2022}, consisting of the United States (US) State-of-the-Union (\Sotu) addresses that were given by the various US presidents each year.  The dataset we obtained contains 233 addresses delivered by 46 presidents, starting with President George Washington's first \Sotu~address in 1790. The last address in our dataset is President Biden's 2022 \Sotu~address. There were a few notable exceptional addresses within the dataset that we note here. \Sotu~addresses can be delivered either in writing or orally, and during the majority of the nineteenth century were delivered in writing. This changed in 1913, and since then, most addresses are given orally. In 1946, President Truman delivered a written address consisting of 25,000 words. In 1981, President Carter also delivered a written address. All addresses were included in our analysis. 
Generally speaking, \Sotu's have a defining place in the US and the world's agenda. Presidents use these opportunities to communicate with the public, set forth their agenda, and make specific policy proposals while addressing the legislative institution~\cite{hoffman2006addressing}. \Sotu~addresses have become a key tool in exercising presidential power and, as such, were the focus of much research, with respect to the topics raised in the addresses and their effect. 

Working with the highly complex textual \Sotu~dataset, which contains well-known topics, events, and meaning, enables us to demonstrate LvS transformation process and the analytical power it provides in a meaningful yet easy-to-understand manner.  

\subsection{Preprocessing} To characterize the \Sotu~dataset using LvS, we consider each year as a time bin, dividing the period from 1790 to 2021 into 232 time bins. The address given each year is then represented as a term frequency distribution vector. Neglecting the order of the words or their grammar, a distribution vector represents the relative frequency of each term in the address. We further remove stop words such as \{the, a, to\} etc. To emphasize topical content, we focused our analysis on nouns and verbs, which reflect the subject matter and actions described in each speech. We removed adjectives and stop words to reduce stylistic variation. Stemming and lemmatization were not applied, in order to retain distinctions that may still carry contextual or rhetorical relevance. Each time bun then consists of the term frequency in the speech. It was then normalized by the length to create a probability distribution vector.  As each year is a time bin, a president who served for four years and gave four addresses will have four probability frequency distribution vectors, one in each corresponding time bin. 
The resulting probability distributions across the timeline are of {\em various lengths}, containing different number of elements.

\subsection{Transformation} 
The first step is to create the \tcr, as depicted in Eq.~(\ref{eq:tcr}). 
The choice to calculate the mean across the time bins rather than to aggregate the frequency of each element all time bins and normalize it gives equal importance to each time bin. Thus, a time bin containing much information can not overshadow other time bins. In the case of \Sotu, a very long address like Truman's 1946 address would still only contribute 1/232 of the weight to the mean of each element.   

The resulting SOTU \tcr~ contains 20,136 terms. The ten most prevalent words across the years are \textit{state, government, year, congress, united, nation, people, country, american,} and \textit{may}. The least frequent word along the timeline is \textit{buffer}.

We then create a time bin Suprisability Profile (SP) for each address, according to Eqs~\ref{eq:sp-init},~\ref{eq:sp-sign},~and~\ref{eq:sp-theta}. At each time bin in the \Sotu~ dataset, the SP represents the terms that most distinguish that year's presidential address, highlighting what is emphasized more and what is emphasized less, or missing, compared with the \tcr.

\subsection{Surprisability Profiles capturing presidential stylistic and topic choices}
Here are a few examples of presidential style and topic choices, as found in the SP. Notable for George Washington is his discussion of ``provisions'', ``appointments'' of ``commissions'', and ``laws'' and ``institutions''. Yet, he seldom references the ``world'', a highly prevalent term in addresses, ranked 14th in the \tcr. 
Woodrow Wilson increased the use of the terms ``action'' and ``necessary'' in 1916, referring to internal matters, while significantly reducing references to ``peace'', ``war'', or the ``world''. This trend was reversed in 1917, a few months before the U.S. joined World War I.  
Similarly, in an abrupt change from Lyndon B. Johnson, Richard Nixon hardly mentions the ``war'' until 1974. Moving to more recent times, an interesting stylistic pattern in Barack Obama's addresses is his frequent use of the word ``know'', although its usage declines in each subsequent year, alongside a decreasing trend in references to the ``government''. Another notable underused term is ``condition'', a relatively common word. Obama rarely uses it, and when he does, he often refers to ``pre-existing conditions''.  
Another interesting stylistic choice is Donald Trump's lack of use of the word ``may'', a highly prevalent term otherwise (ranked tenth in the \tcr). Trump used the term twice in his first address in 2017 but did not use it at all in his subsequent speeches during that presidential term. Notable in his speeches is his frequent repetition of guest names.  

To validate that the SPs capture important information about the presidents' styles, we identify similarities between presidents based on their surprising terms in their addresses and cluster them accordingly. We then compare the resulting clustering to a well-established reference classification.

\begin{figure*}[!ht]
  \centering
  \includegraphics[width=.96\linewidth]{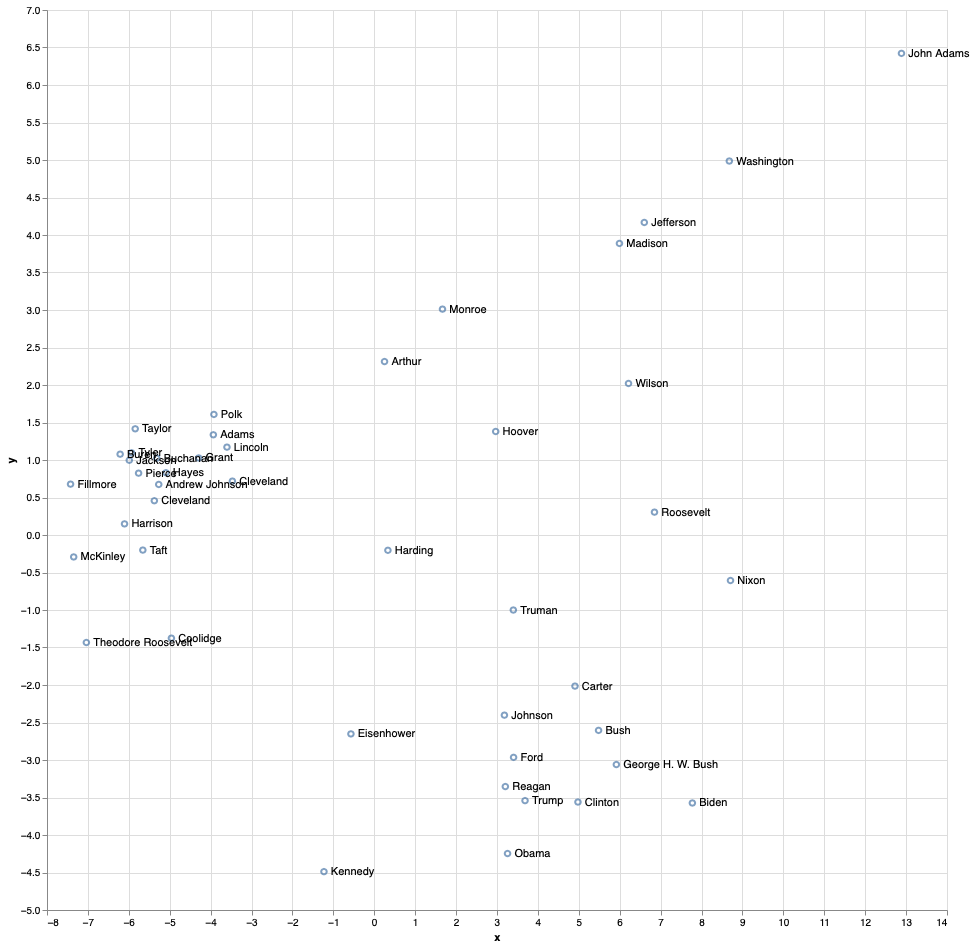}
  \caption{Two-dimensional PCA representation of the differences across LvS's transformation of presidential addresses. PCA1 (X-axis) accounts for 67\% of the variance, while PCA2 (Y-axis) accounts for 6.8\%. Pairwise surprisal distances between State of the Union addresses, with notable years annotated. The highlighted 1942 speech stands out due to a multivariate shift in language. The associated surprisal profile (right) shows increased use of war-related terms (war, fighting, enemy, hitler, victory) and decreased use of institutional language (government, public, law), indicating a broad rhetorical change reflective of World War II.}
  \label{Sotu-pca}
\end{figure*}

Figure~\ref{Sotu-pca} presents a dimensionality reduction representation of a distance matrix computed for all presidents. To construct this matrix, each \Sotu~ address given by a president was compared to all \Sotu~addresses given by other presidents using the Manhattan distance between their SPs. The average of these distances was then calculated for each presidential pair. Additionally, we computed intra-president distances to measure how consistent a president’s SPs are across their own speeches.
To validate our results, we reference the work of Jacques Savoy~\cite{savoy2015text}, who analyzed presidential speech similarities using text clustering. In a similar approach, Savoy applied PCA to a distance matrix computed between the presidents' speeches, focusing on part-of-speech usage. Their analysis contrasted the frequency of determiners and prepositions against pronouns, modal verbs, and adverbs. However, in their case, the two principal components accounted for 40.3\% and 15.9\% of the variance, while in our case, they account for 67\% and 6.8\%, respectively.

Our findings align with Savoy’s in several aspects. Like us, they identify a distinct cluster on the left of Figure~\ref{Sotu-pca}, including Presidents McKinley, Taft, Cleveland (twice), Hayes, Pierce, Harrison, Jackson, Theodore Roosevelt, Fillmore, and Coolidge. However, while we place President Hoover further from this group, Savoy positions him closer, though they also find no clear stylistic matches for his speeches.

Similarly, both analyses place Presidents Eisenhower, Kennedy, and Ford in the same region. However, our findings regarding Presidents Bush (both senior and junior), Reagan, Clinton, and Obama differ from Savoy’s. That said, both studies find that Presidents Carter and Bush Sr. have relatively similar speech patterns. Notably, Savoy's analysis does not include Presidents Trump and Biden.

In addition to the modern and post-war clusters, a third cluster emerged, comprising early presidents such as Washington, Jefferson, and Madison. This group is distinguished by both stylistic and topical differences that align with the historical context of their presidencies. These speeches tend to avoid nation-centric terms like \{America, Americans, or people\}, which appear more frequently in later eras. Instead, they emphasize themes relevant to the early republic, such as \{debt, militia, friendship, and rivers\}. For example, Washington references the ``militia'' far more than any other president, Jefferson focuses on ``vessels'' and ``rivers'', and Madison frequently discusses the ``British'' and ``war''. These patterns reflect the geopolitical concerns and rhetorical style of the time and demonstrate the capacity of LvS to capture historically grounded distinctions in both topic and emphasis.

Overall, this comparison reinforces the validity of our method, demonstrating its ability to effectively capture stylistic patterns in presidential addresses.

\subsection{Interpretable anomaly detection in \Sotu~Addresses over time}

\begin{figure*}[!ht]
  \includegraphics[width=.8\linewidth]{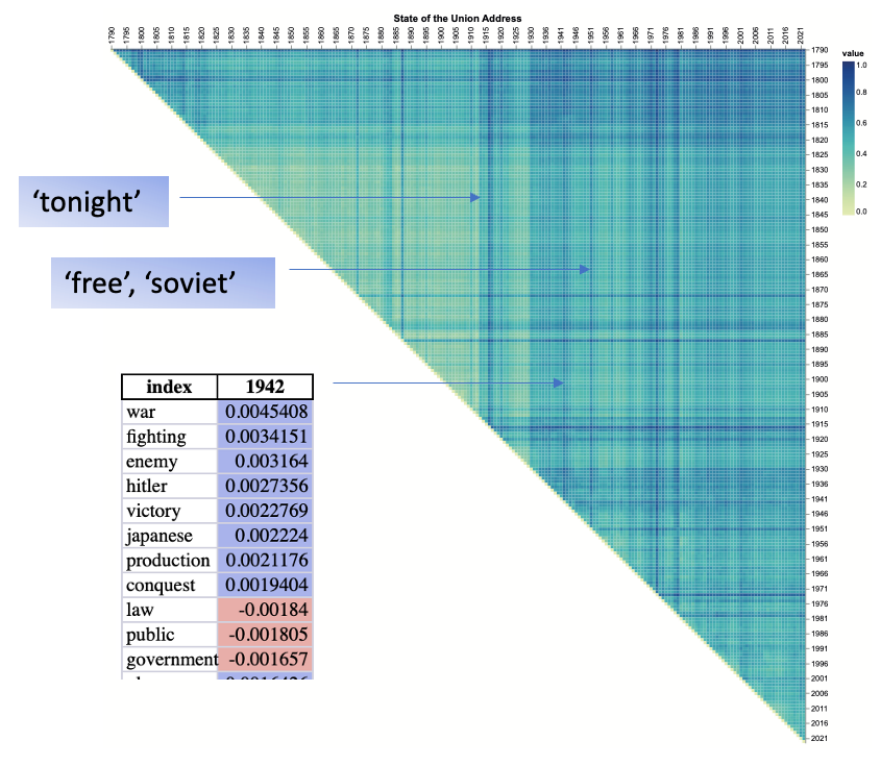}
 \caption{\textbf{Comparison of SPs found in \Sotu~ addresses across all years. Values were normalized for readability. Annotations highlight notable points of change: the top surprising term in 1913; the ten most surprising terms in the 1942 \Sotu~ address—with overused terms in blue and underused terms in pink—showing a broad, multivariate shift toward war-related language (war, fighting, enemy, victory) and away from institutional terms (government, public); and the top two surprising terms in 1951 (free, soviet).} }
 \label{Sotu-all}
\end{figure*}
Figure~\ref{Sotu-all} provides a visual anomaly detection of \Sotu~ addresses over the years, highlighting points where certain terms significantly deviate from the norm. By normalizing the values, the figure makes it easier to spot moments of linguistic or thematic change in presidential rhetoric.

Several notable patterns emerge in the data. A clear linguistic shift occurred when the U.S. entered World War I, but even earlier, in 1913, Woodrow Wilson introduced a significant change. The most surprising term of that year, tonight, reflects a major transition—before 1913, \Sotu~ addresses were delivered in writing, whereas from that year onward, they were generally presented orally.

While all speeches produce multivariate surprisal profiles, we focus on the 1942 address as a representative example due to space constraints. As shown in Figure~\ref{Sotu-all}, the 1942 speech exhibits a coordinated shift across multiple features: increased use of war-related terms (war, fighting, enemy, hitler, victory) and decreased emphasis on institutional language (government, public, law). This pattern reflects a broader rhetorical realignment rather than a spike in any single term, demonstrating LvS’s capacity to surface interpretable, multivariate anomalies across time.

In 1951, six months after the Korean War began, Harry S. Truman delivered a speech marked by two key surprising terms: \textit{free} and \textit{Soviet}. These shifts in language reflect how presidential rhetoric adapts to major historical events, making them clear points of linguistic change.

\subsubsection{Highly Variable \Sotu~Terms Across the Years}
\label{sec:variable}
We present here the fluctuations of a few selected highly variable terms over time. These terms are among the most variable in \Sotu~addresses and were chosen because they exhibit distinct trends, highlighting different patterns of change in presidential rhetoric and thematic emphasis.
\begin{figure*}[!ht]
    \centering
    \includegraphics[width=0.9\linewidth]{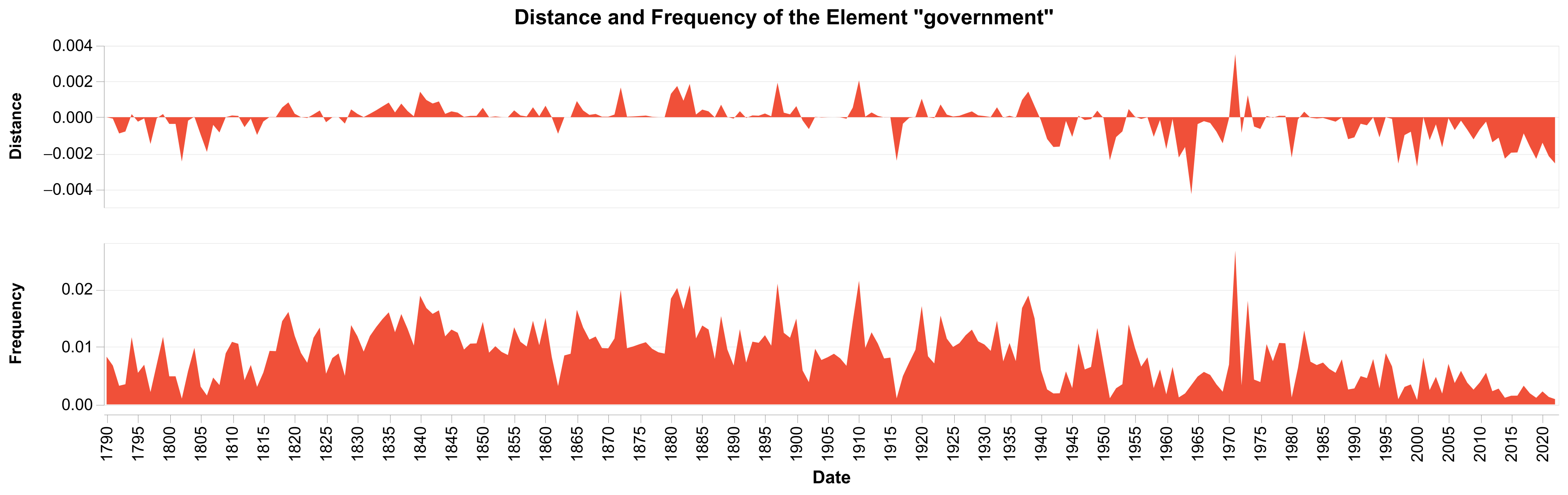}
    \caption{Dynamics of the term \textit{government}. The upper panel shows the variability by displaying the JSD from the \tcr, while the lower panel shows the actual frequency along the timeline.}
    \label{fig:gov}
\end{figure*}

\begin{figure*}[!ht]
    \centering
    \includegraphics[width=0.9\linewidth]{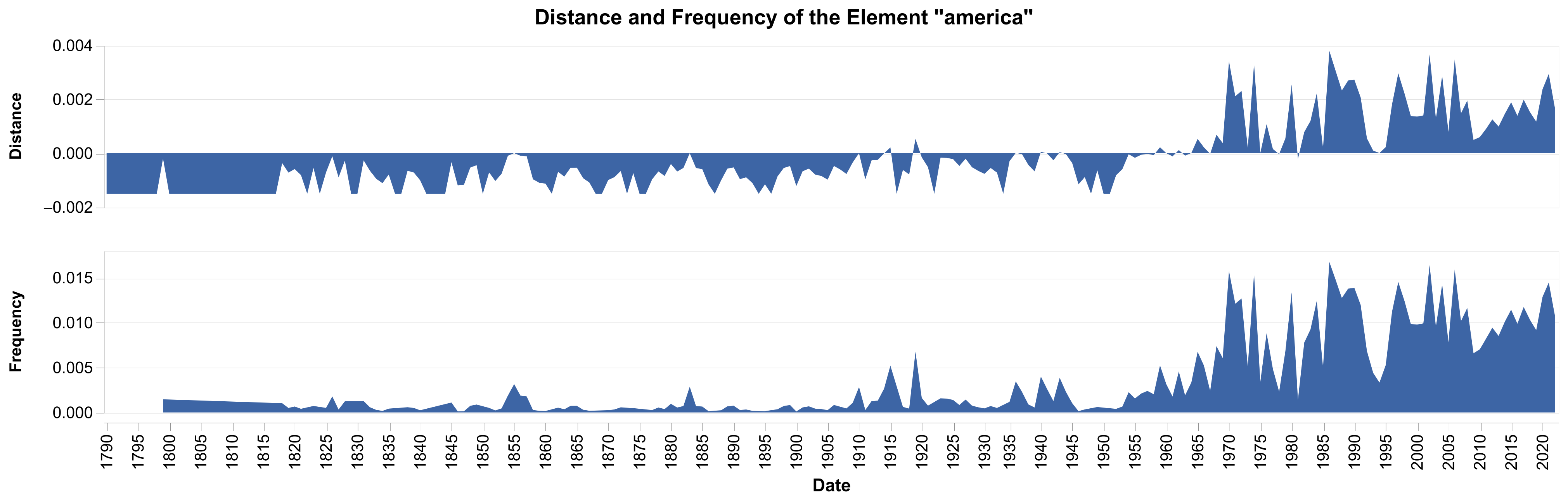}
    \caption{Dynamics of the term \textit{america}. The upper panel shows the variability by displaying the JSD from the \tcr, while the lower panel shows the actual frequency along the timeline.}
    \label{fig:amer}
\end{figure*}

\begin{figure*}[!ht]
    \centering
    \includegraphics[width=0.9\linewidth]{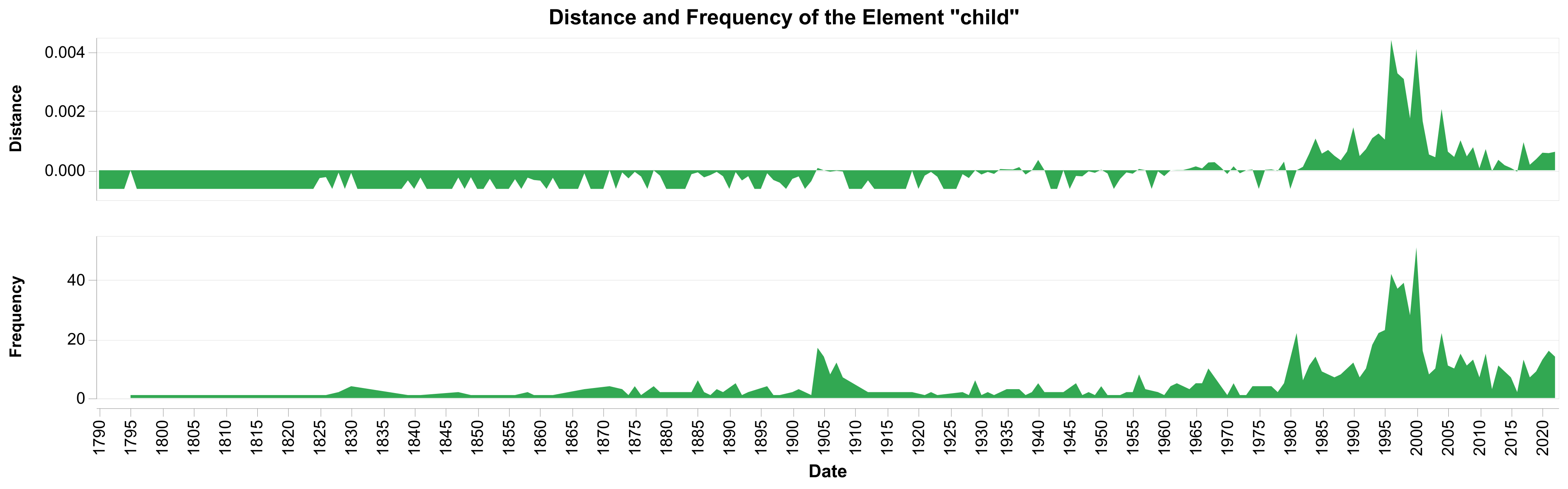}
    \caption{Dynamics of the term \textit{child}. The upper panel shows the variability by displaying the JSD from the \tcr, while the lower panel shows the actual frequency along the timeline.}
    \label{fig:child}
\end{figure*}

Figures~\ref{fig:gov},~\ref{fig:amer}, and~\ref{fig:child} illustrate how the usage of key terms—\textit{government, America}, and \textit{child}—has changed over time in \Sotu~addresses, capturing both their frequency and variability. Each figure consists of two panels: the upper panel represents the JSD from the \tcr, highlighting how much the term's usage deviates from expected, while the lower panel shows its actual frequency across the years.

Examining Figure~\ref{fig:gov}, which tracks the term \textit{government}, we see that it was frequently mentioned throughout the 19th and early 20th centuries, peaking in importance at various points before declining significantly after 1980. The variability of this term also shows noticeable shifts, suggesting that while the word remained a key part of presidential rhetoric for a long time, its role in political discourse has changed, especially in modern addresses.

In Figure~\ref{fig:amer}, the term \textit{America} shows a gradual increase in both frequency and variability over time. Its presence in speeches was relatively low in the 19th century but saw a substantial rise from the mid-20th century onward, particularly during moments of national significance such as World War II, the Cold War, and post-9/11 addresses. This suggests a growing emphasis on national identity and unity in presidential rhetoric.

Similarly, Figure~\ref{fig:child} shows that the term \textit{child} was rarely mentioned before the 20th century, with noticeable peaks emerging in the 1960s and 1990s, suggesting that child welfare became a more prominent topic during this period, particularly peaking in President Clinton's addresses. The increasing variability in later years indicates that while the term appears more frequently, its contextual significance may be evolving.

Overall, these figures highlight how LvS transformation can be used presidential rhetoric adapts to historical and political contexts, with some terms gaining prominence over time while others decline in usage. The variability patterns suggest that certain words, even when frequently used, can carry different connotations and importance depending on the era in which they appear.

\section{Discussion and Conclusion}
\label{sec:discussion}
Here, we introduced Learning via Surprisability (LvS) as a novel transformation approach, which is useful in characterizing high-dimensional time-series data through a representation of the timeline (i.e., a temporally ordered dataset) through Timeline Center Representation (\tcr) and of the time bins through their Surprisability Profiles (SP). The SP of a time bin includes the most surprising elements relative to the expected \tcr. LvS transformation provides a novel approach for analyzing high-dimensional timeline data, effectively addressing challenges such as high dimensionality, complex distributions, and sparsity. By drawing inspiration from cognitive science and the concept of surprisal, LvS identifies the most unexpected elements in a dataset, offering a powerful and interpretable method for anomaly detection and transformation. Unlike existing methods that either reduce dimensionality at the expense of interpretability or rely on predefined heuristics, LvS preserves context in the timeline by emphasizing features that deviate the most from expected patterns. \review{LvS is applied in an offline setting, where the full timeline is available for constructing a reference distribution and computing surprisal profiles.}

The ability of LvS to detect anomalies and highlight significant deviations was demonstrated in multiple real-world datasets. In the SWaT dataset, where the goal was to detect cyber-physical anomalies in an industrial water treatment system, LvS performed on par with or better than state-of-the-art unsupervised anomaly detection methods such as Isolation Forest, Local Outlier Factor (LOF), and One-Class SVM. While conventional models struggled to find a balance between precision and recall, LvS consistently identified attack instances with high accuracy and recall. The approach of comparing each time bin to a timeline-wide expectation proved highly effective in flagging anomalies, which in this case corresponded to cyberattacks. The interpretability of LvS adds another layer of utility—rather than simply labeling an observation as an anomaly, it provides insight into the features responsible for the deviation. This is a key advantage over black-box machine learning models, which often struggle to provide explanations for their decisions.

In addition to anomaly detection capabilities, LvS also excels at capturing contextual shifts over time. In the analysis of mortality causes worldwide, the transformation enabled a clear visualization of trends, outliers, and abrupt changes in global health data spanning three decades. While traditional statistical approaches focus on overall trends, LvS highlighted specific moments in time when major shifts occurred, such as the spike in deaths due to conflict in 1994, corresponding to the Rwandan genocide, or the unusual increase in mortality caused by natural disasters in 2004, 2008, and 2010. The ability to pinpoint such changes and, more importantly, to interpret the features responsible for them makes LvS a valuable tool for time-series analysis.

A similar effect was observed in the textual domain, where LvS was applied to the State of the Union addresses spanning more than two centuries. By transforming each speech into a surprisal-based representation, the method revealed stylistic shifts, policy priorities, and linguistic trends across different presidencies. For example, the sudden rise of war-related terms in 1917, before the U.S. entered World War I, and the dominance of military and conflict-related words in Roosevelt’s 1942 address at the height of World War II were captured clearly through the transformation. At the same time, LvS identified less obvious patterns, such as the declining use of the word ``government'' over the last few decades or the unique stylistic tendencies of individual presidents. The clustering of presidents based on their surprisal profiles showed strong alignment with independent linguistic studies, reinforcing the validity of the transformation in capturing meaningful distinctions in writing style and thematic focus.

The LvS transformation stands in contrast with existing transformation and anomaly detection techniques, offering a distinctive approach that models changes over time rather than aggregating entity distributions into a single population-wide representation, as seen in Latent Personal Analysis \cite{mokryn2021domain}. Unlike Latent Personal Analysis, which provides a structured global representation of entities by capturing deviations from a global reference distribution, LvS dynamically constructs an evolving expectation of the timeline, enabling the detection of temporal shifts and anomalies. Similarly, while classical transformation methods such as PCA and Fourier analysis offer effective dimensionality reduction, they often fail to preserve local interpretability, particularly in non-stationary datasets. Fourier transform and its variants are widely used for time-series analysis but are primarily designed for frequency decomposition rather than anomaly detection, limiting their ability to identify contextual outliers in high-dimensional datasets.

Beyond transformation methods, LvS also diverges from traditional and contemporary anomaly detection techniques. Classical statistical anomaly detection approaches, such as Gaussian mixture models~\cite{li2016anomaly}, kernel density estimation~\cite{silverman2018density}, and Mahalanobis distance-based outlier detection~\cite{thudumu2020comprehensive}, rely on assumptions about the underlying data distribution, making them less effective in high-dimensional or sparse datasets where distributions are complex and non-Gaussian. More recent machine learning-based approaches, including deep autoencoders~\cite{zhou2017anomaly}, generative adversarial networks (GANs) for anomaly detection~\cite{zenati2018efficient}, and one-class SVMs~\cite{oza2018one}, offer state-of-the-art performance in identifying anomalies but often lack interpretability. These models typically function as black boxes, providing little insight into why a given observation is classified as anomalous. Moreover, methods such as Isolation Forest and LOF assess the degree of isolation of individual data points, but their effectiveness diminishes in high-dimensional spaces due to the curse of dimensionality~\cite{verleysen2005curse}. Unlike these approaches, LvS retains both computational efficiency and interpretability by identifying the most surprising features that differentiate each time bin from the timeline’s overall expectation, allowing users to understand and contextualize anomalies rather than treating them as abstract deviations.

Despite its advantages, LvS has several limitations that should be acknowledged. \review{First, the method relies on the assumption that past distributions accurately represent expected behaviors, which may be challenged in the presence of concept drift, where the underlying data distribution changes over time~\cite{limit_1_1}.} To address this, future work could integrate adaptive learning mechanisms that dynamically update the baseline distribution. One potential approach is to incorporate online learning models that continuously adjust the TCR in response to new data, ensuring the method remains sensitive to evolving patterns~\cite{limit_2_1,limit_2_2}. Second, LvS is inherently retrospective, making it unsuitable for real-time anomaly detection without further adaptations. Third, the current implementation of LvS uses a fixed surprisal threshold ($\theta$) to determine which elements are included in each time bin’s surprisability profile. While this approach provides useful and interpretable representations, it is heuristic and may not generalize optimally across datasets or domains. Future work could explore adaptive thresholding methods based on the statistical distribution of surprisability scores, cumulative contribution to the total divergence (e.g., an elbow point), or data-driven optimization techniques. Such approaches could improve the stability and generalizability of LvS representations, especially in dynamic or noisy settings. Fourth, LvS currently relies on fixed temporal binning. The choice of bin size significantly influences the method’s sensitivity: small bins may lead to high variance in distributions due to limited sample size, while large bins can obscure transient or time-sensitive anomalies. This trade-off becomes particularly challenging in datasets that are temporally sparse or exhibit oscillatory behavior, where identifying stable reference behavior and choosing an appropriate temporal resolution are nontrivial. One possible direction to address this challenge is the use of Multi-Scale Entropy (MSE)~\cite{humeau2015multiscale}, which captures regularities across different time scales and may offer a way to mitigate bin-size sensitivity. Additionally, Scaled Bregman Divergence, proposed by Wang et al. (2025)~\cite{wang2025anomaly}, may serve as a useful extension to LvS. This divergence generalizes to non-overlapping supports and satisfies metric properties—potentially improving robustness in diverse data conditions. Exploring such alternatives offers a promising avenue for future development.

In summary, LvS provides a novel framework for transforming and analyzing high-dimensional timeline data, effectively bridging the gap between classical transformation techniques and modern anomaly detection methods. Its ability to highlight significant shifts while preserving contextual meaning makes it a promising tool for a wide range of applications, from cybersecurity monitoring and financial risk assessment to epidemiological studies and linguistic research. Future directions include applications in more specialized fields, such as biomedical signal processing or fraud detection in financial transactions.

\section*{Acknowledgments} During the early stages of conceptualizing this project, we benefited greatly from thoughtful discussions with Petal Mokryn, Uri Hershberg, and Alex Abbey.  We are also grateful to Alex Abbey for his additional support during the coding process.

\section*{Declarations}

\subsection*{Funding}
This study received no funding. 

\subsection*{Conflicts of Interest}
The authors declare no conflict of interest.

\subsection*{Code and Data Availability}
\review{The code used for all experiments, along with the processed CoD and SOTU datasets, is publicly available at \url{https://github.com/ScanLab-ossi/LvS}. The SWaT dataset is available through the link provided in the manuscript.}
\subsection*{Author Contribution}
\textbf{Osnat Mokryn:} Conceptualization, Methodology, Formal analysis, Investigation, Data Curation, Writing - Original Draft, Writing - Review \& Editing, Visualization, Supervision, Project administration. \\
\textbf{Teddy Lazebnik: }Validation, Investigation, Data Curation, Writing - Original Draft, Writing - Review \& Editing. \\
\textbf{Hagit Ben Shoshan: }Software, Formal analysis, Investigation, Data Curation,  Visualization.

\section*{References}
\appendix
\section*{Appendix A: Hyperparameter Search Spaces}
The following hyperparameter search spaces were used in the grid search procedure to tune all models for anomaly detection in the SWAT dataset, as described in Section~\ref{sec:swat}. The hyperparameter search spaces for each algorithm were as follows:

\subsection*{Isolation Forest}
\begin{itemize}
    \item \texttt{n\_estimators}: [100, 200, 300]
    \item \texttt{max\_samples}: ['auto', 0.5, 0.7]
    \item \texttt{contamination}: [0.01, 0.05, 0.1]
    \item \texttt{max\_features}: [0.5, 0.7, 1.0]
\end{itemize}

\subsection*{SVM (One-Class SVM)}
\begin{itemize}
    \item \texttt{kernel}: ['rbf', 'linear', 'poly', 'sigmoid']
    \item \texttt{nu}: [0.01, 0.05, 0.1, 0.25]
    \item \texttt{gamma}: ['scale', 'auto', 0.1, 1]
\end{itemize}

\subsection*{LOF (Local Outlier Factor)}
\begin{itemize}
    \item \texttt{n\_neighbors}: [5, 10, 20, 50]
    \item \texttt{contamination}: [0.01, 0.05, 0.1]
    \item \texttt{algorithm}: ['auto', 'ball\_tree', 'kd\_tree', 'brute']
\end{itemize}

\subsection*{AE (Autoencoder)}
\begin{itemize}
    \item \texttt{epochs}: [50, 100, 150]
    \item \texttt{batch\_size}: [32, 64, 128]
    \item \texttt{learning\_rate}: [0.001, 0.0001]
    \item \texttt{latent\_dim}: [5, 10, 20]
\end{itemize}

\subsection*{LSTM}
\begin{itemize}
    \item \texttt{epochs}: [10, 20, 30]
    \item \texttt{batch\_size}: [16, 32, 64]
    \item \texttt{learning\_rate}: [0.001, 0.0001]
    \item \texttt{hidden\_units}: [32, 64]
    \item \texttt{sequence\_length}: [10, 20, 30]
\end{itemize}

\end{document}